\documentclass[prl,reprint,amsmath,amssymb,aps,preprintnumbers,showpacs]{revtex4-1}
\usepackage[colorlinks=true,linkcolor=black, citecolor=black,urlcolor=black]{hyperref} 
\usepackage{amstext,amsmath,amssymb,amsfonts,mathrsfs}   
\usepackage{graphicx}  
\usepackage{braket} 
\usepackage{bbm} 
\usepackage[british,UKenglish,USenglish]{babel}
\usepackage[dvipsnames]{xcolor}
 \usepackage{tikz}
 \usetikzlibrary{plotmarks,calc,decorations, decorations.pathmorphing, patterns,hobby}
 \usetikzlibrary{arrows}
 \tikzstyle dynkin node=[very thick,shape=circle,draw,inner sep=0pt,minimum size=5mm]
 \tikzstyle dynkin line=[very thick]
 \tikzstyle inverse line=[gray,line width=1.46pt,line cap=round, dash pattern=on 0pt off 2\pgflinewidth]
 \tikzstyle red phase=[red,decoration={snake,amplitude=0.1mm,segment length=1.6mm},decorate]
 \tikzstyle blue phase=[blue,decoration={snake,amplitude=0.1mm,segment length=0.9mm},decorate]
 \tikzstyle green phase=[green,decoration={snake,amplitude=0.1mm,segment length=0.9mm},decorate]
 \tikzstyle brown phase=[brown,decoration={snake,amplitude=0.1mm,segment length=0.9mm},decorate]
 
 \usetikzlibrary{decorations.pathmorphing}
 \tikzstyle arrow=[thick,rounded corners=18pt,-latex]
 \tikzstyle box=[draw,rounded corners,outer sep=4pt]
\tikzstyle B node=[outer sep=0pt]
\tikzstyle Q node=[inner sep=1pt,outer sep=0pt]

\definecolor{purple_nice}{rgb}{0.4,0.2,0.7}
\definecolor{fuel_blue}{RGB}{42,162,185}
\definecolor{YInMn_blue}{RGB}{46, 80, 144}
\definecolor{ultramarine}{RGB}{63, 0, 255}
\definecolor{KLEIN_blue}{rgb}{0, 0.18, 0.65}

\newcommand{\de}{\text{d}}

\newcommand{\TTbar}{T\overline{T}}
\newcommand{\mir}[1]{\tilde{#1}}

\newcommand{\cdd}{\chi}

\newcommand{\parL}{\boldsymbol\xi}
\newcommand{\parR}{\boldsymbol\eta}
\newcommand{\fields}{\boldsymbol\Phi}

\newcommand{\partic}{\text{p}}
\newcommand{\hole}{\text{h}}

\newcommand{\defo}{\alpha}
\newcommand{\pardef}{\boldsymbol\alpha}
\newcommand{\chargerev}{\check{\mathbf{J}}}

\begin{document}

\vspace*{0cm}

\title{Flow equations for generalised \texorpdfstring{$\TTbar$}{T-Tbar} deformations}
\author{Guzm\'an Hern\'andez-Chifflet$^{1}$}
\email{guzmanhc@fing.edu.uy}
\author{Stefano Negro$^{2}$}
\email{stefano.negro@stonybrook.edu}
\author{Alessandro Sfondrini$^{3,4,5}$}
\email{alessandro.sfondrini@unipd.it}

\affiliation{${}^1$  Instituto de F\'isica, Facultad de Ingenier\'ia, Universidad de la Rep\'ublica, Montevideo, 11300, Uruguay\\
${}^2$  C.~N.~Yang Institute for Theoretical Physics, Stony Brook University, Stony Brook, NY 11794, USA\\
${}^3$ Institut f\"ur theoretische Physik, ETH Z\"urich, Wolfgang-Pauli-Stra{\ss}e 27, 8093 Z\"urich,
Switzerland\\
${}^4$ Dipartimento di Fisica e Astronomia ``Galileo Galilei'', Universit\`a degli Studi di Padova, via Marzolo 8, 35131 Padova, Italy\\
${}^5$ Istituto Nazionale di Fisica Nucleare, Sezione di Padova, via Marzolo 8, 35131 Padova, Italy}
\date{\today}

\begin{abstract}
\noindent
We consider the most general set of integrable deformations extending the $\TTbar$ deformation of two-dimensional relativistic QFTs. They are CDD deformations of the theory's factorised S-matrix related to the higher-spin conserved charges. Using a mirror version of the generalised Gibbs ensemble, we write down the finite-volume expectation value of the higher-spin charges, and derive a generalised flow equation that every charge must obey under a generalised $\TTbar$ deformation. This also reproduces the known flow equations on the nose.
\end{abstract}


\pacs{02.30.Ik, 11.30.Ly, 11.55.Ds.}
\maketitle

\paragraph{Introduction.}
Our understanding of physics has been unfailingly advanced by the study of exactly solvable models --- from the Kepler problem to the latest advances in interacting quantum field theories (QFTs). A powerful illustration of this approach is given by integrable QFTs (IQFTs) in two spacetime dimensions, see \textit{e.g.}~\cite{Dorey:1996gd,Bombardelli:2016rwb} for reviews.
Physically, we can think of IQFTs as arising from deforming a two-dimensional conformal field theory (CFT) by carefully chosen relevant operators, inducing a renormalisation group flow. The resulting theory is not conformal, but is nonetheless endowed with infinitely-many independent mutually-commuting conserved quantities --- a remnant of conformal symmetry~\cite{Bazhanov:1994ft,Bazhanov:1998dq,Bazhanov:1996dr,Negro:2016yuu}. This  constrains the dynamics to the point that it allows to efficiently compute a wealth of observables --- something very remarkable for an interacting QFT!

Given an exactly-solvable theory it is natural to ask how much we may modify it while preserving its solvability. Recently we started to realise that deforming CFTs (or QFTs) by \emph{irrelevant} operators might be as physically interesting as the better-understood relevant deformations. Moreover, this paves the way to quantitatively describe a new class of theories.
The prime example of irrelevant deformations is the ``$\TTbar$'' deformation~\cite{Smirnov:2016lqw,Cavaglia:2016oda}, built out of the stress-energy tensor~$T_{\mu\nu}$~\cite{Zamolodchikov:2004ce}. This arises by infinitesimally deforming the Hamiltonian density~$H$ by the composite operator $O_{\TTbar}=T^{0\mu}T^{1\nu}\epsilon_{\mu\nu}$ and integrating the resulting flow, $\partial_\alpha H=O_{\TTbar}$. $\TTbar$-deformed theories are remarkable, and despite intensive study still mysterious: on the one hand, they can be related to two-dimensional gravity~\cite{Dubovsky:2017cnj,Dubovsky:2018bmo,Conti:2018tca,Ishii:2019uwk}, or to random geometries~\cite{Cardy:2018sdv}. On the other hand, they can be also reformulated in terms of string theory~\cite{Cavaglia:2016oda,Baggio:2018gct, Frolov:2019nrr,Hashimoto:2019wct, Sfondrini:2019smd,Callebaut:2019omt,Tolley:2019nmm} (see also \cite{Dubovsky:2012wk,Caselle:2013dra} for earlier observations of the relation between strings and $\TTbar$) and holography~\cite{McGough:2016lol,Giveon:2017nie,Chakraborty:2019mdf} and even defined for spin chains~\cite{Pozsgay:2019ekd,Marchetto:2019yyt}. 
$\TTbar$ deformations are special as they preserve many symmetries: supersymmetry~\cite{Baggio:2018rpv,Chang:2018dge, Jiang:2019hux,Chang:2019kiu}, modular invariance~\cite{Aharony:2018bad}, and most remarkably \textit{integrability}~\cite{Smirnov:2016lqw,Cavaglia:2016oda}. By this we mean that if the original theory is a CFT, or an IQFT, its infinitely-many conserved charges are preserved by the deformation. More is true: even if the original theory \emph{is not integrable}, the deformation is exactly solvable: the finite-volume spectrum $\{H_n\}$ of $\TTbar$-deformed theories obeys the Burgers equation
\begin{equation}
\label{eq:Burgers}
    \partial_\alpha H_n(R,\alpha) = H_n(R,\alpha)\partial_RH_n(R,\alpha) + \frac{1}{R}P_n{}^2\,,
\end{equation}
where $P_n=2\pi N_n/R$, $N_n\in\mathbb{Z}$, is the  momentum and $H(R,0)$ the original Hamiltonian. Similar equations may be written for the $\TTbar$ deformation of more general charges~\cite{LeFloch:2019wlf}.
Still, $\TTbar$ is just one of \emph{infinitely many} similar integrable deformations of relativistic QFTs~\cite{Smirnov:2016lqw}. This letter investigates such arbitrary deformations and derives the analogue of the flow equation~\eqref{eq:Burgers} for generic observables --- not just energy and momentum.

To do so, we firstly review how $\TTbar$ deformations and their generalisations are defined in terms of the S-matrix of any IQFT~\cite{Mussardo:1999aj}. This particular formulation makes it possible to employ integrability techniques such as the thermodynamic Bethe ansatz (TBA)~\cite{Yang:1968rm,Zamolodchikov:1989cf} to derive~\eqref{eq:Burgers} and to study the theory~\cite{Cavaglia:2016oda}. For generalised deformations the TBA will not suffice. Ordinarily,~\eqref{eq:Burgers} tells us that tuning~$\alpha$ corresponds to changing~$R$; we will see that more general deformations correspond to changing new parameters, cousins of~$R$, that may be interpreted as \emph{twists} of the fields' boundary conditions. Periodic boundary conditions $\fields(0)=\fields(R)$ will then be modified conjugating the right-hand side by an additional unitary operator~$e^{iJ\eta}$, where $\eta\in\mathbb{R}$ is the twist. We will see that such twists may described using the generalised Gibbs ensemble (GGE)~\cite{PhysRevLett.98.050405} for a \emph{mirror} theory, which we will introduce in the sense of Refs.~\cite{Ambjorn:2005wa,Arutyunov:2007tc}; then $\eta$ plays the role of a \emph{chemical potential}. This \textit{mirror} GGE construction is to our knowledge new, though work in this direction appeared earlier in~\cite{Bajnok:2004jd, Arutyunov:2010gu,Ahn:2011xq,Bajnok:2019mpp}. With this machinery we derive the analogue of~\eqref{eq:Burgers} for an infinite family of integrable deformations --- our main result~\eqref{eq:flow_eqs}.

\paragraph{The factorised S-matrix.}
Due to the existence of IQFT conservation laws, scattering is heavily constrained: the only allowed processes are sequences of elastic two-particle collisions. Hence all scattering amplitudes may be written in terms of the two-to-two particle S-matrix~$S_{12}$, whose matrix structure must satisfy a consistency condition, the celebrated Yang-Baxter equation~\cite{Zamolodchikov:1978xm}.
This, along with global symmetries, unitarity, analyticity and crossing symmetry, constrains~$S_{12}$. Often $S_{12}$ is almost entirely determined by these requirements --- it can be \emph{bootstrapped}~\cite{Zamolodchikov:1978xm}.
The solution is not unique, however: it is only defined up to a Castillejo-Dalitz-Dyson (CDD) factor~\cite{Castillejo:1955ed}.

\paragraph{CDD deformations.}
A two-dimensional relativistic S-matrix is most easily described by introducing the \emph{rapidity} $\theta$, related to energy and momentum as $H=m\cosh\theta$ and $p=m\sinh\theta$, where $m$ is the mass. Then $S_{12}$ depends on the difference of rapidities $S_{12}=S(\theta_1-\theta_2)$, and each  S-matrix entry is meromorphic on the $\theta$-plane. Linearly-realised symmetries and the Yang-Baxter equation leave an overall prefactor $\cdd(\theta)$ undetermined. If $S_{12}$ is appropriately normalised, $\cdd(\theta)$ is a meromorphic function on the complex plane satisfying Hermitian analyticity~\cite{Miramontes:1999gd}, $\cdd(\theta^*)^*=\cdd(\theta)$, unitarity, $\cdd(\theta)\cdd(-\theta)=1$, and crossing symmetry, $\cdd(\theta)=\cdd(i\pi-\theta)$.
This means that we may set $\cdd(\theta)=e^{i\Sigma(\theta)}$ where $\Sigma(\theta)$ is a $2\pi i$-periodic meromorphic, real-analytic function. The space of such $\chi$s defines a family of integrable theories, at least in terms of their S-matrix. There are two natural ways of parametrising $\chi(\theta)$. We can define it by its poles and resonances, $\chi(\theta|\mathbf{a})=\prod_j \tanh(\theta-ia_j)/2$: such singularities have a clear physical interpretation in terms of the infrared properties of theory. Otherwise we can write a Fourier series,
\begin{equation}
\label{eq:irrelevantdefo}
    \Sigma(\theta|\pardef)=\sum_{j\ \text{odd}} \defo_j\, e^{-j\theta}\,,\quad
    \defo_j=-\defo_{-j}\,,\quad
    \alpha_j\in\mathbb{R}\,,
\end{equation}
where we restricted the coefficients using unitarity, real analyticity and crossing.
Each $\alpha_j$ in~\eqref{eq:irrelevantdefo} affects the large-$\theta$ asymptotic of~$S(\theta)$, corresponding to an integrable \emph{irrelevant} deformation \cite{Smirnov:2016lqw,Zamolodchikov:talk,Rosenhaus:2019utc}. In particular,  $\Sigma(\theta)=\alpha m^2\sinh\theta$  yields the $\TTbar$ flow~\eqref{eq:Burgers}~\cite{Cavaglia:2016oda,Dubovsky:2017cnj}. More general deformations correspond to composite operators of the form $J_{(j)}^\mu J_{(-j)}^\nu\epsilon_{\mu\nu}$~\cite{Smirnov:2016lqw}, where $J_{(j)}^\mu$ are the infinitely-many conserved currents of the integrable theory --- the higher-spin currents~\footnote{%
In complex coordinates we may write, for  $j>0$, $J_{(j)}^\mu=(T_{j+1},\Theta_{j-1})$ and $J_{(-j)}^\mu=(\overline{\Theta}_{j-1},\overline{T}_{j+1})$.}. The charges $J_{j}$ commute among themselves and act diagonally on multi-particle scattering states, \textit{i.e.}\ on states where particles are well-separated:
\begin{equation}
\label{eq:chargeaction}
    J_j |\theta_1,\dots,\theta_N\rangle
    =
    \sum_{k=1}^N J_j(\theta_k)\,|\theta_1,\dots,\theta_N\rangle\,,
\end{equation}
with $J_j(\theta)=e^{j\theta}$.
We shall describe the finite-volume properties of such CDD deformations.

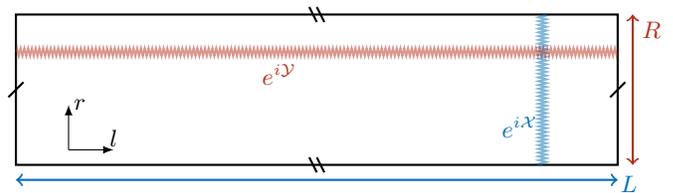
\begin{figure}
\centering
\begin{tikzpicture}
\draw[thick] (-4,2) rectangle (4,0);
\draw[thick] (-3.9,1.1) -- (-4.1,0.9);
\draw[thick] (4.1,1.1) -- (3.9,0.9);
\draw[thick] (-0.1,2.1) -- (-0,1.9);
\draw[thick] (0,2.1) -- (0.1,1.9);
\draw[thick] (-0.1,0.1) -- (-0,-0.1);
\draw[thick] (0,0.1) -- (0.1,-0.1);
\draw[thick,BrickRed,<->] (4.2,0) -- (4.2,2);
\node[] at (4.45,1.8) {$\color{BrickRed}R$};
\draw[thick,NavyBlue,<->] (-4,-0.2) -- (4,-0.2);
\node[] at (4.15,-0.25) {$\color{NavyBlue}L$};
\draw[black,-latex] (-3.3,0.2) -- (-3.3,0.8);
\node[] at (-3.15,0.8) {$r$};
\draw[black,-latex] (-3.3,0.2) -- (-2.7,0.2);
\node[] at (-2.7,0.35) {$l$};
\draw[thick,NavyBlue, opacity=0.5,decoration = {zigzag,segment length = 0.5mm, amplitude = 0.5mm},decorate] (3,0) -- (3,2);
\node[] at (2.7,0.5) {$\color{NavyBlue}e^{i\mathcal{X}}$};
\draw[thick,BrickRed,opacity=0.5,decoration = {zigzag,segment length = 0.5mm, amplitude = 0.5mm},decorate] (-4,1.5) -- (4,1.5);
\node[] at (-0.5,1.2) {$\color{BrickRed}e^{i\mathcal{Y}}$};
\end{tikzpicture}
 \caption{\label{fig:torus} An Euclidean theory on a torus. The Cartesian coordinates $(l,r)$ are periodically identified, $l\cong l+L$ and $r\cong r+R$. Later we shall twist the boundary conditions of fields $\fields(l,r)$ by $\mathcal{X}$ and $\mathcal{Y}$. 
 In the direct theory, $\sigma\equiv r$ so that $\mathcal{X}$ is related to a charge (integrated over space) while $\mathcal{Y}$ leads a twist along the spatial direction, which may be interpreted as a defect. In the mirror, this is reversed.}
 \end{figure}

\paragraph{Finite-volume (and finite-temperature) theories.}
Consider a two-dimensional Euclidean theory, defined on a torus like in Figure~\ref{fig:torus}. We take one radius to be very large, $L\gg R$, and eventually $L\to\infty$. There are two ways to obtain Minkowski theories. Firstly, we may Wick-rotate and define
\begin{equation}
\label{eq:direct}
\text{direct theory}:\qquad
\sigma \equiv r\,,\quad
\tau \equiv il\,.
\end{equation}
We call this the \emph{direct theory}; it lives in finite-volume~$R$ but at almost-zero temperature~$1/L$. Conversely, we may set
\begin{equation}
\label{eq:mirror}
\text{mirror theory}:\qquad
\mir{\sigma} \equiv l\,,\quad
\mir{\tau} \equiv ir\,.
\end{equation}
This is the \emph{mirror theory}, at finite temperature~$1/R$ but in large volume; we denote mirror quantities with tildes. For simplicity we consider relativistic integrable theories with one particle flavour, so that the two-particle S-matrix is a \emph{function}~$S(\theta)$.
Given (\ref{eq:direct}--\ref{eq:mirror}), we may go from the direct theory to its mirror by 
\begin{equation}
H\to i\mir{p}\,,\qquad
p\to -i\mir{H}\,,
\end{equation}
Hence, up to a parity transformation, the mirror theory is the analytic continuation of the direct one by $\mir{\theta}\equiv\theta-\tfrac{i\pi}{2}$ (half of a crossing transformation). This leaves $S_{12}$ and the dispersion unchanged. This construction yields an equality between the \textit{thermal} partition function of the mirror model $\text{Tr}[e^{-R\mir{H}}]$ and finite-volume spectrum of the direct one $\sum_n e^{-L H^{(n)}}$: when $L\to\infty$ the mirror free-energy density $\mir{F}(R)$ and direct-theory ground state energy are related as $R\mir{F}(R) = H^{(0)}(R)$, see~\cite{vanTongeren:2016hhc} for a recent review.
We will consider such quantities in presence of  boundary conditions \emph{twisted} by charges~$\mathcal{X}$ and $\mathcal{Y}$ as in Figure~\ref{fig:torus}. These twists break Poincar\'e invariance, but leave local properties such as dispersion and S-matrix unaffected.

\paragraph{Mirror generalised Gibbs ensemble (GGE).}
To study the finite-volume direct theory, we compute the twisted (generalised) free energy of the \emph{mirror} theory. Since $mL\gg1$, the mirror Bethe-Yang equations are approximately correct~\cite{Luscher:1985dn,Luscher:1986pf}. The $L$-cycle twist  affects the spatial boundary conditions for the mirror theory (Figure~\ref{fig:torus}). This means replacing the monodromy $e^{i\mir{p}(\mir{\theta})L}\equiv e^{H(\mir{\theta}+i\pi/2)L}$ by $e^{X(\mir{\theta}+i\pi/2|\parL)L}$, which depends on a set of parameters~$\parL=\{\xi_j\}$ that identify the charges appearing in $\mathcal{X}$: $X(\theta|\parL)=\sum_j\xi_j J_j(\theta)$. Then the Bethe-Yang equations are
\begin{equation}
\label{eq:BYeq}
X(\mir{\theta}_k+i\tfrac{\pi}{2}|\parL)\,L+\sum_{l\neq k}^N\log S(\mir{\theta}_k-\mir{\theta}_{l})=2\pi in_k\,,
\end{equation}
with $n_k\in\mathbb{Z}$.
As $X(\theta)$ is defined in the \textit{direct theory}, in~\eqref{eq:BYeq} we need to analytically continue it to the real-mirror line; all terms in the equation are purely imaginary for $\mir{\theta}_k\in\mathbb{R}$.
In the thermodynamic limit $N\sim mL\gg1$,
\begin{equation}
\label{eq:logBYeq}
\varrho_{\partic}(\mir{\theta})+\varrho_{\hole}(\mir{\theta})=\frac{1}{2\pi i}\partial_{\mir{\theta}} X(\mir{\theta}+i\tfrac{\pi}{2}|\parL)+[\varphi * \varrho_{\partic}](\mir{\theta})\,,
\end{equation}
in terms of the densities of particles and holes $\varrho_{\partic},\varrho_{\hole}$ and of the Kernel $\varphi_{12}=\partial_{1} \log S_{12}/(2\pi i) $, all on the real-mirror line
\footnote{We define the convolutions on the real-mirror line $\varphi*f=\int\de z \varphi_{1z}f_z $ and $f*\varphi=\int\de z f_z \varphi_{z2}$.}.
To twist the $R$-cycle we introduce chemical potentials in
\begin{equation}
\label{eq:miror_partfun}
Z(L,R|\parL,\parR) = \text{Tr}
\exp\big[-R \tilde{Y}(\parR)
\big]\Big|_{\Phi_{\mathcal{X}(\parL)}}\,.
\end{equation}
Here $\tilde{Y}$ is the operator in the mirror theory corresponding to the charge $\mathcal{Y}$,  parametrized as $R\tilde{Y} = \sum_{j} \eta_j\,\tilde{J}_j$. Each $\tilde{J}_j$ acts diagonally on \emph{mirror} states~$|\tilde{\theta}_1,\dots, \tilde{\theta}_N\rangle$, \textit{cf.}~\eqref{eq:chargeaction}.
Here $\mir{Y}$ is an operator on the mirror-theory Hilbert space.  We extremise the free energy  in terms of $\varrho_{\partic},\varrho_{\hole}$. Standard manipulations (see \textit{e.g.}~\cite{vanTongeren:2016hhc}) yield the mirror GGE equations
\begin{equation}
\label{eq:GGEeq}
\varepsilon(\mir{\theta}|\parR) = R \mir{Y}(\mir{\theta}|\parR) + [\Lambda(\parR)*\varphi]\,(\mir{\theta})\,,
\end{equation}
where $\Lambda(\mir{\theta}|\parR)=\pm\log(1\mp e^{-\varepsilon(\mir{\theta}|\parR)})$
in terms of the mirror-pseudoenergy $\varepsilon(\mir{\theta}|\parR)$ for bosons and fermions, respectively~%
\footnote{The pseudoenergy satisfies $e^{\varepsilon}=\pm 1+(\varrho_{\partic}+\varrho_{\hole})/\varrho_{\partic}$ for bosons and fermions, respectively.}. This equation is real on the real-mirror line.
Remarkably, this equation is sensitive to $\parR$ but \emph{not to} $\parL$. The mirror free-energy density depends on both parameters:
\begin{equation}
\label{eq:generealisedmirrorF}
R\tilde{F}(R|\parL,\parR) =\frac{1}{2\pi i} \int\text{d}\mir{\theta}\; \partial_{\mir{\theta}}{X}(\mir{\theta}+i\tfrac{\pi}{2}|\parL)\,\Lambda(\mir{\theta}|\parR)\,.
\end{equation}
Here $X$ is the direct-theory {operator} corresponding to the Euclidean $\mathcal{X}$, parametrised as $X = \sum_j\xi_jJ_j$, \textit{cf.}~\eqref{eq:chargeaction}. In~\eqref{eq:generealisedmirrorF} $X$ is on the \emph{mirror line}~\footnote{For $\parL=0$, $X=H\to i\mir{p}$.}.
For the direct theory, we have
\begin{equation}
Z(L,R|\parL,\parR) = \text{Tr}
\exp\big[-L X
\big]\Big|_{\Phi_{\mathcal{Y}}}\,,
\label{eq:mirror_partfun}
\end{equation}
Thus for $L\to\infty$ we relate~\eqref{eq:generealisedmirrorF} to the direct theory as
\begin{equation}
\label{eq:generaliseddirectGS}
R\tilde{F}(R|\parL,\parR) = \sum_{j}\xi_j J_{j}^{(0)}(R|\parR)\,.
\end{equation}
By comparing the $\xi_j$-dependence of (\ref{eq:generealisedmirrorF}--\ref{eq:generaliseddirectGS}) we find
\begin{equation}
\label{eq:LGGEvalue}
J_j^{(0)}(R|\parR)=\frac{1}{2\pi i}
\int\text{d}\mir{\theta}\;
\partial_{\mir{\theta}}{J}_j(\mir{\theta}+i\tfrac{\pi}{2})\,\Lambda(\mir{\theta}|\parR)\,.
\end{equation}
Note that we expressed the vacuum value~$J_j^{(0)}$ of the \emph{direct-theory charge}~$J_j$ (on a state with spatial boundary conditions twisted by~$Y$) through a mirror-theory integral. The relations between defects and chemical potentials in the direct and mirror theories were previously investigated in \cite{Bajnok:2004jd,Ahn:2011xq} .

\paragraph{Excited states.}
Our derivation may also be extended to excited states of the direct theory. They should be described by the same equations with integrals on some state-dependent contour~$\Gamma$~\cite{Dorey:1996re} rather than on the real mirror line.  (See~\cite{Bajnok:2019mpp,Borsi:2019tim,Pozsgay:2019xak} for recent investigations of  excited-state expectation values.) The equations then differ by residues picked up between $\Gamma$ and the real-mirror line at points~$\theta_k$ where~$e^{-\varepsilon(\theta_k)}=\mp 1$, as the log becomes singular. This may happen when $\theta_k$ is on (or around) the real line \emph{of the direct theory} (hence the lack of tilde). Integrating by parts the GGE equations (\ref{eq:GGEeq},~\ref{eq:LGGEvalue}) we find residues of the form $\log S(\mir{\theta}-\theta_k)$ and $J_j(\theta_k)$, respectively. This modifies the vacuum equations by driving terms. In particular in~\eqref{eq:LGGEvalue} the driving term is $\sum_k J_j(\theta_k)$, where the charge~$J_j$ is evaluated \textit{in the direct theory} owing to analytic continuation to~$\theta_k$. Remark than when~$mR\gg1$ the GGE should reduce to the asymptotic result; indeed in this limit $\sum_k J_j(\theta_k)$ dominates and reproduces the asymptotic eigenvalue of $J_j$ on a well-separated direct-theory state, \textit{cf.}~\eqref{eq:chargeaction}. 

\paragraph{CDD deformations in the GGE.}
Knowing the finite-volume spectrum with twisted boundary conditions, we can study general CDD deformations of the form~\eqref{eq:irrelevantdefo}.
Such modifications shift linearly the Kernel $\varphi(\theta_1-\theta_2)=\varphi(\mir{\theta}_1-\mir{\theta}_2)$. We get
$\varphi(\mir{\theta}_{12})\to \varphi(\mir{\theta}_{12})+
    \frac{1}{2\pi}\sum_{j}j\alpha_{-j}\, e^{j\mir{\theta}_{12}}$.
Then the GGE equation~\eqref{eq:GGEeq} becomes
\begin{widetext}
\begin{equation}
\label{eq:modifiedGGE}
    \varepsilon(\mir{\theta}|\parR,\pardef) =\, R \mir{Y}(\mir{\theta}|\parR) + \big[\Lambda(\parR,\pardef)*\varphi\big](\mir{\theta}) +\!\! \sum_{j\ \text{odd}} \frac{\defo_{j}}{i^{j+1}} e^{j\mir{\theta}} J_{-j}(\parR,\pardef)\,,
    \quad
    J_j(\parR,\pardef)=
    \frac{1}{2\pi i}\int\text{d}\mir{\theta}\,
    \partial_{\mir{\theta}} e^{j(\mir{\theta}+i\tfrac{\pi}{2})}\,\Lambda(\mir{\theta}|\parR,\pardef).
\end{equation}
\end{widetext}
By comparing this with~\eqref{eq:LGGEvalue}, we see that~$J_j(\parR,\pardef)$ is the ground-state value of a \emph{direct-theory charge} with density $J_j(\theta)=e^{j\theta}$, \textit{i.e.}\ of the direct-theory higher-spin charges~\eqref{eq:chargeaction}.
We can simplify the GGE by setting all $\xi_j=0$, as this identification also works for infinitesimal~$\xi_j$s.
We see that the new term in~\eqref{eq:modifiedGGE} can be reabsorbed into~$\mir{Y}(\tilde{\theta}|\parR)$ by a constant (but charge-dependent) shift of the parameters $\eta_j$, namely $\eta_j\to\eta_j+i^{j+1}\alpha_jJ_{-j}$, implying
\begin{equation}
\label{eq:Funct_rel_pseudoen}
     \varepsilon(\mir{\theta}|\parR,\pardef)
     =
      \varepsilon(\mir{\theta}|\parR+\pardef\chargerev,\mathbf{0})\,,
\end{equation}
where $\chargerev$ is the ordered set $\{i^{j+1}J_{-j}\}$. Hence \textit{all physical quantities} derived from the GGE will depend on $(\parR+\pardef\chargerev)$ only.

\paragraph{The generalised flow equation.} A consequence of~\eqref{eq:Funct_rel_pseudoen} is that every conserved charge satisfies a flow equation. A charge~$J_j$, like all quantities computed from the pseudoenergy, obeys $J_{j}(\parR,\pardef) = J_j(\parR+\pardef\chargerev(\parR,\pardef),\mathbf{0})$. Hence, defining the differential operator
\begin{equation}
\label{eq:differential}
    \mathsf D_n = i^{n-1}\frac{\partial}{\partial \alpha_n} + J_{-n}(\parR,\pardef)\frac{\partial}{\partial \eta_n} + J_{n}(\parR,\pardef)\frac{\partial}{\partial \eta_{-n}}\,,
\end{equation}
for any \emph{positive} odd integer $n$, we obtain by a direct computation
\begin{equation}
\label{eq:linearsystem}
    \mathsf D_n J_j(\parL,\pardef)=
    \sum_{\ell\ \text{odd}} M_{j\ell}(\parR+\pardef\chargerev)\, \mathsf D_n J_{-\ell}(\parL,\pardef)\,,
\end{equation}
where $M_{j\ell}(\textbf{z})=\frac{\ell}{\vert\ell\vert}i^{\vert\ell\vert+1}\defo_{\vert\ell\vert}\tfrac{\partial}{\partial z_\ell}J_{j}(\textbf{z},\mathbf{0})$.
Therefore, as long as the operator with matrix elements $M_{j\ell}+\delta_{j\ell}$ is non-singular (which is the case for small deformations), the only solution to~\eqref{eq:linearsystem} is
\begin{equation}
    \mathsf D_n J_j(\parL,\pardef)=0\,,
\label{eq:flow_eqs}
\end{equation}
for all positive odd integers $n$ and all odd integers $j$. This gives an infinite set of flow equations obeyed by \emph{every physical observable} $\mathcal{O}(\parR,\pardef)$ derived from the GGE: $\mathsf D_k \mathcal{O}(\parR,\pardef) = 0$, since $J_j$ is a basis for any such~$\mathcal{O}$.

\paragraph{Recovering the Burgers equation.}
The flow equation~\eqref{eq:Burgers} of ordinary $\TTbar$ follows from \eqref{eq:flow_eqs} when we have only $\alpha_1 \equiv \tfrac{1}{2}\alpha$. 
The only non-vanishing chemical potentials are $\eta_{\pm 1} \equiv \tfrac{1}{2}m e^{\pm \zeta}$. Here $\zeta$ is an auxiliary parameter (a chemical potential for the direct-theory energy), useful~\cite{Cavaglia:2016oda} to derive the inhomogenous Burgers equation; we  will eventually set~$\zeta=0$.
Note that $\zeta$  enters the GGE~\eqref{eq:GGEeq} as a rapidity shift, 
\begin{equation}
\label{eq:theta0shift}
\varepsilon(\tilde{\theta}\vert R,\zeta;\alpha) = \varepsilon(\tilde{\theta}+\zeta\vert R,0;\alpha)\,.    
\end{equation}
The physical chemical potential is the mass $m$. As the GGE depends on $m$ through the dimensionless combination $mR$, we may trade $\partial_m$ for $\partial_R$. 
There is only one flow operator, $\mathsf D \equiv \frac{1}{2}\mathsf D_1$,
\begin{equation}
    \mathsf D =\frac{\partial}{\partial\alpha} + H(R,0;\alpha)\frac{\partial}{\partial R} - \frac{1}{R} P(R,0;\alpha)\frac{\partial}{\partial \zeta} \,.
\end{equation}
Here we expressed $\partial_{\parR}$ as $\partial_R=\partial_m$ and $\partial_{\zeta}$, and we introduced the total energy and momentum $2H = J_1 + J_{-1}$ and $2P = J_1 - J_{-1}$. Let us now compute the $\partial_{\zeta}$ derivatives.
Note first that using~\eqref{eq:theta0shift} and shifting the integration measure in~\eqref{eq:LGGEvalue}, we find that $J_j(R,\zeta;\alpha) = e^{-j\zeta} J_j(R,0;\alpha)$. Therefore, omitting the arguments for convenience, $\partial_{\zeta} H = - P$ and $\partial_{\zeta}P = -H$. Hence our flow equation $\mathsf  D H\vert_{\zeta = 0} = 0$ is precisely the Burgers equation \eqref{eq:Burgers}. The other equation, $\mathsf D  P\vert_{\zeta = 0} = 0$, gives that $\partial_\alpha P=0$ if we also use that $\partial_{R}P=-P/R$ (which can be derived from the GGE equations).
This is expected from the quantisation of $P$.
Repeating this argument for $\mathsf D J_j=0$ reproduces the $\TTbar$ flow equations for $J_j$ proposed in~\cite{LeFloch:2019wlf}.

\paragraph{Conclusions and outlook.}
We derived flow equations~\eqref{eq:flow_eqs} for generalised $\TTbar$ deformations that constrain all the GGE observables. We argued this for the vacuum, but clearly our starting point~\eqref{eq:Funct_rel_pseudoen} holds for excited states too --- these are governed by the same GGE equations~(\ref{eq:GGEeq},~\ref{eq:LGGEvalue}) up to changing the integration contour. Hence~\eqref{eq:flow_eqs} is completely general. Our construction uses relativistic invariance sparingly, so that it should be possible to extend it to non-relativistic setups
like those of~\cite{Arutyunov:2004vx,Arutyunov:2009ga} and~\cite{Guica:2017lia,LeFloch:2019wlf,Conti:2019dxg}.
Finally, this mirror GGE might be useful beyond the present case, to study twists in relativistic and non-relativistic integrable models.

It would be interesting to study the generalised flows for some simple systems. For a supersymmetric free theory (with Neveu-Schwarz conditions) the GGE trivialises (much like in~\cite{Baggio:2018gct,Dei:2018mfl}) and we only have to deal with algebraic equations. The ground-state GGE of a single-flavour theory can also be studied relatively easily. Either case would require numerical investigations, though. A preliminary analysis points to qualitative difference to~$\TTbar$. This is expected as in~\eqref{eq:modifiedGGE} even a tiny generalised deformation yields the dominant contribution to the pseudoenergy at large-$|\mir{\theta}|$, and dramatically affects the convergence properties of the GGE integrals. We plan to report on this elsewhere~\cite{GHer_SNeg_ASfo_19}. 

In \cite{Aharony:2018bad} it was found that~\eqref{eq:Burgers} is the only flow equation for finite-volume energy levels preserving modular invariance. We should investigate whether generalised deformations preserve modular covariance of GGE partition functions and whether this requirement uniquely defines them. Another important question is whether these deformations can be obtained by introducing gauge fields coupled to the higher-spin currents of the IQFTs, similarly to how $\TTbar$ may be obtained by coupling the undeformed theory to a gravitational sector.

Finally, in \cite{Negro:2013wga} (see also \cite{Negro:2013rxa,Negro:2017xwc}), the GGE was proposed as a tool to access, in the specific case of sinh-Gordon model, the finite-volume expectation values of local operators. We expect this perspective to be useful to investigate the expectation values of the deformed theories presented in this letter.

\begin{acknowledgments}
\paragraph{Acknowledgments.}
We are grateful to Andrea Dei, Sergey Frolov and Roberto Tateo for comments on a draft of this manuscript, and to Sergei Dubovsky for useful discussions. G.H.\ would like to thank Nicol\'as Wschebor and Miguel Campiglia for related discussions. S.N.\ is grateful to Z.\ Komargodski, M.\ Mezei and A.\ Zamolodchikov, for stimulating and insightful discussions, and to NYU for the kind hospitality.
We would like to thank the participants of the workshop \textit{New frontiers of integrable deformations} for stimulating discussions that contributed to these results.
S.N.'s work is funded by NSF Award PHY-1620628, and by the FCT Project PTDC/MAT-PUR/30234/2017 ``Irregular connections on algebraic curves and Quantum Field Theory''. A.S.'s work was funded by ETH Career Seed Grant No.~SEED-23 19-1, as well as by the NCCR SwissMAP, funded by the Swiss National Science Foundation. 
\end{acknowledgments}

\bibliographystyle{apsrev4-1}
\bibliography{refs}

\end{document}